\documentclass[runningheads,a4paper]{llncs}

\usepackage{amsmath}
\usepackage{amssymb}
\usepackage{graphicx}
\usepackage[linesnumbered,algoruled]{algorithm2e}
\usepackage{subfigure}

\usepackage{url}
\urldef{\mailsa}\path|Fangfang.Li@student.uts.edu.au |
\urldef{\mailsb}\path|{Guandong.Xu,Longbing.Cao}@uts.edu.au|
\urldef{\mailsc}\path|{Yanchang.Zhao,Klaus.Felsche}@immi.gov.au|

\begin{document}

\mainmatter  

\title{Coupling Analysis Between Twitter and Call Centre}

\titlerunning{Coupling Analysis}

%
%

\author{Fangfang Li$^{1,2}$
\and Yanchang Zhao$^2$ \and Klaus Felsche $^2$ \and Guandong Xu$^1$\and Longbing Cao$^1$}
\authorrunning{F. Li, Y. Zhao, K. Felsche, G. Xu, L. Cao}
\institute{$^1$Advanced Analytics Institute, University of Technology, Sydney, Australia \\
\mailsa \\
\mailsb\\
$^2$ Department of Immigration and Border Protection, ACT, Australia\\
\mailsc
}

\maketitle

\begin{abstract}
Social media has been contributing many research areas such as data mining, recommender systems, time series analysis, etc. However, there are not many successful applications regarding social media in government agencies. In fact, lots of governments have social media accounts such as twitter and facebook. More and more customers are likely to communicate with governments on social media, causing massive external social media data for governments. This external data would be beneficial for analysing behaviours and real needs of the customers. Besides this, most governments also have a call centre to help customers solve their problems. It is not difficult to imagine that the enquiries on external social media and internal call centre may have some coupling relationships. The couplings could be helpful for studying customers' intent and allocating government's limited resources for better service. In this paper, we mainly focus on analysing the coupling relations between internal call centre and external public media using time series analysis methods for Australia Department of Immigration and Border Protection. The discovered couplings demonstrate that call centre and public media indeed have correlations, which are significant for understanding customers' behaviours.
\end{abstract}

\section{Introduction}

Social media has been attracting more and more attention from many research areas such as data mining, text mining, recommender systems\cite{DBLP:YangSL12_circleRS} and etc. The related topics in social media mainly include: event discovery, group or community discovery, social network analysis, personalized recommendation, sentiment analysis and so on. However, regarding the applications of social media analysis, there is little work in government agencies, actually even the mature techniques have not been well applied in governments such as the Australia Department of Immigration and Border Protection (DIBP).

Much helpful information such as immigration or visa-related policies has been posted by DIBP's accounts on twitter. The people who are of interest to Australia may pay attention to DIBP's tweets. For example, some users may reply to the tweets by asking further questions or just expressing a positive or negative opinion, others may repost the tweets to propagate the information to their friends. We believe that the study and analysis of the social media will potentially improve and better understand the business of the governments. In addition to communicate with DIBP on social media websites, many people are likely to directly ring DIBP's call centre to ask their questions or check the progress of visa applications. This internal call centre data is very helpful for discovering people's interests on DIBP and allocating the limited resources for better public service. The internal call centre data may also have some couplings with the external social media data. For example, during one period, the call centre may receive many queries about visa applications. If we check the social media websites, maybe there are many visa-related queries as well. When Australia commences a new policy, probably there will be many discussions regarding this policy on social media websites, the calls regarding the policy will be also increasing in the following days or weeks. This motivates us to meditate the following questions: (1) Whether the internal call centre and external social media have correlations? (2) If yes, is there some time lag between the two platforms? (3) How the significant events related to DIBP influence the internal call centre and external social media and their correlations? Based on the internal call centre data, DIBP has many predictive models such as risk evaluation, intent analysis. We believe that the consideration of external data and the correlations with internal call centre data would contribute to improving the performance of predictive or monitoring models.

To analyse the couplings between call centre and public media, we first give a toy example in table \ref{tab:toyExample_callCentre} to illustrate how the data look like. The essential attributes of every phone call includes ``ID", ``Date", ``Duration", ``Disposition Code" and ``Disposition Code Text". By these attributes, the exact time and the duration of the phone call are recorded, after the interaction, DIBP's call centre agent will categorize this enquiry into different topics by ``Disposition Code" and the detail attribute ``Disposition Code Text". Similarly on social media, every tweet also has many attributes such as ``tweetID", ``tweetText", ``postTime", and ``userName". The attributes of tweets does  not contain the disposition code for categorization, but we can use text mining techniques to determine the topic information based on ``tweetText" attribute.

\begin{table}[htbp]
  \centering
  \caption{A Toy Example for Call Centre}
    \begin{tabular}{|c|c|c|c|c|}
      \hline
      ID & Date & Duration & Disposition Code & Disposition Code Text \\
      ID1 & 6/12/2013 & 1min & 457 visa & 457 visa application progress \\
      ID2 & 6/12/2013 & 10 min & student visa & 574 visa \\
      ID3 & 6/12/2013 & 3 min & 457 visa & 457 visa application progress \\
      ID4 & 7/12/2013 & 15 min & 600 visit & 600 visit visa \\
      ID5 & 7/12/2013 & 1min & 600 visit & 600 electronic visa \\
      ID6 & 7/12/2013 & 5min & skilled migration & skilled selection \\
      \hline
    \end{tabular}
    \label{tab:toyExample_callCentre}
\end{table}

Because both of the data sets contain the ``time" information, we could transform the data sets into time series, then apply time series analysis methods to study the coupling relations. The contributions of the paper are concluded as follow:
\begin{itemize}
\item We automatically extracted DIBP's external twitter data using R programming language;
\item We investigated the characteristics of DIBP's internal call centre data by time series analysis;
\item We analysed the correlations between internal call centre and social media by a symbolic representation time series analysis method;
\item We conducted experiments to verify the couplings between the internal and external data.
\end{itemize}

The rest of the paper is organized as follows. Section 2 presents the related work. In Section 3, we propose and detail the coupling analysis framework for call centre and public media. Experimental analysis and some discovered couplings are presented in Section 4 followed by the conclusion.

\section{Related Work}

With the development of social network, more and more researchers have been trying to incorporate social media into relevant research areas such as sentiment analysis\cite{DBLP:conf/icwsm/OConnorBRS10}, time series analysis\cite{DBLP:conf/dexa/ZhuXC12}, recommender systems\cite{DBLP:MaZLLK11_SocialRegularization}, text mining\cite{yanchangZhao} etc. Regarding time series analysis, the main research topics include: time series index and search\cite{Hochheiser:2004:DQT:993176.993177}\cite{Gunopulos:2001:TSS:376284.375808}\cite{DBLP:journals/kais/CamerraSPRK14}, clustering\cite{DBLP:journals/kais/KeoghL05}\cite{DBLP:conf/pakdd/LinVKGLYL05}\cite{DBLP:conf/edbt/LinVKG04}, classification\cite{Sejdic:2009:TFR:1465726.1465790}\cite{DBLP:conf/icdm/00010ZUK13}\cite{DBLP:conf/sdm/Chen0K13}, summarization, anomaly detection and so on. For example, through some similarity or dissimilarity measures\cite{DBLP:journals/datamine/BatistaKTS14}\cite{DBLP:journals/datamine/WangMDTSK13}, we can query relevant time series, or partition similar time series into a group by clustering or classification methods. However, there are not many successful applications to study the couplings between internal data and external social media data. This strongly motivates us to analyse the couplings using time series analysis techniques.

Time series changing trends are often affected by some seasonal patterns such as cultural festivals like Christmas or Chinese New Year, to better understand the time series, it is often helpful to remove the seasonal patterns which can conceal the true underlying movements in the series. The main methods for time series decomposition \cite{ts_decomposition} include: moving average, additive decomposition, multiplicative decomposition, ARIMA decomposition. In this paper, the simple and mature additive decomposition method implemented in R is applied to remove the seasonal patterns.

\section{Coupling Analysis Framework}

\begin{figure}
\centerline{\includegraphics[scale=0.6]{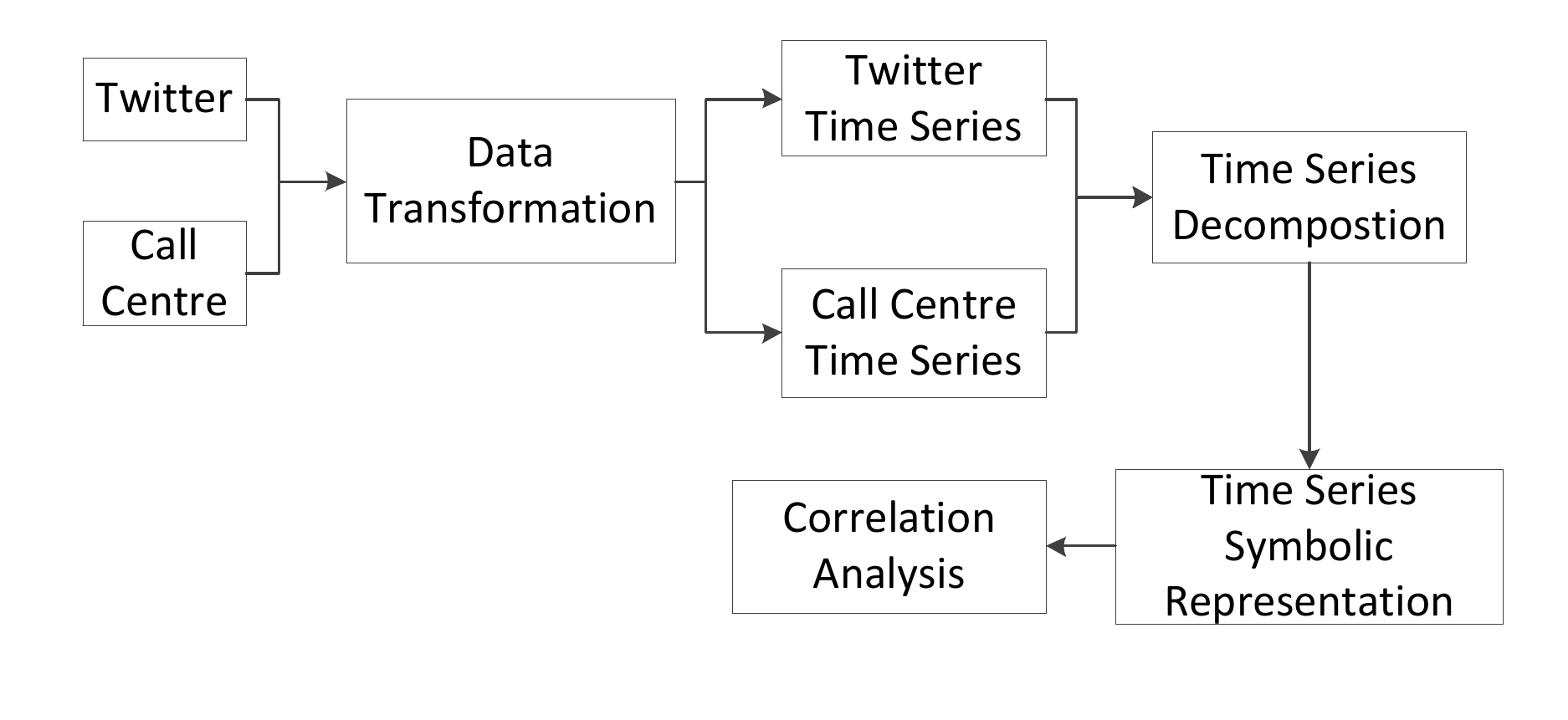}}
\caption{Coupling Analysis for Call Centre and Twitter}
\label{fig_tsCouplingAnalysis_Framework}
\end{figure}

The framework for correlation analysis in this paper is shown in Fig. \ref{fig_tsCouplingAnalysis_Framework}. The main steps include:
\begin{itemize}
  \item Extracting external twitter data;
  \item Integrating different call centre data into specific topics using merging rules;
  \item Transforming external and internal data into time series;
  \item Applying time series decomposition on call centre data to remove seasonal patterns;
  \item Computing correlations similarities and lagged time between call centre and public media;
  \item Using symbolic representation based weekly aggregation on time series for analysing the correlations ;
  \item Extracting interesting changing patterns based on computed similarities and symbolic changing trends.
\end{itemize}

In this section, we introduce the main techniques in this framework.

\subsection{Call Centre Data Integration}
\label{callCentreIntegration}

\begin{figure}
\centerline{\includegraphics[scale=1.0]{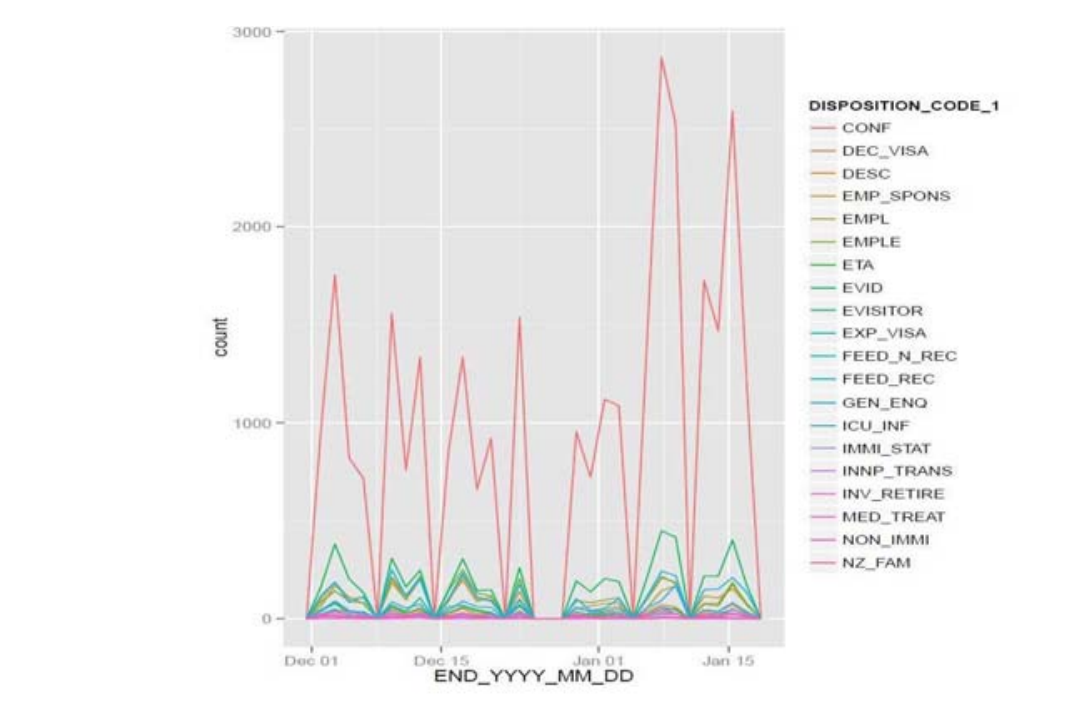}}
\caption{Example of Call Centre in Canberra}
\label{fig_callCentreExample}
\end{figure}

DIBP has a mainland call centre in Canberra and international call centres in London and Ottawa, the categorization labels such as ``Disposition Code" and ``Disposition Code Text" are handled by different staff from different countries, having different file formats. Fig. \ref{fig_callCentreExample} is an example for Canberra's call centre data. It is reasonable to integrate different call centre data together. Therefore, we reorganize the categorization levels and group all the enquiry records into six general categories: ¡°citizen¡±, ¡°work¡±, ¡°study¡±, ¡°visit¡±, ¡°permanent¡± and ¡°other¡± by the following merging rules:

\begin{itemize}
\item Rule 1: $eStudent | student | 560 | 570 | 571 | 572 | 573 | 574 | 575 | 576 | temp\  grad \rightarrow study$
\item Rule 2: $600|676|e600| e651| e676| eta| transit| business \  visitor| eVisitor| visitor|\\sponsor \  visitor|medical \  treatment|carer \rightarrow visit $
\item Rule 3: $400| 417| 456| 457| 462| e400| e417| e457| e462| temp \  long| temp \  short| whm\\ \rightarrow work$
\item Rule 4: $gsm|family \  migration|professional \  migration|partner \  migration|\\rrv|adoption|adult \  migrant \  English \  program|business \  skills|cers|employer \  \\sponsored| employees| employers| NZ \  Family \  Relationship|parent| refugee|\\ remaining \  relative| skilled \  migration| skill \  select|family \  and \  living \rightarrow permanent$
\item Rule 5: $citizenship|conferral|declaratory \  visa|descent|renunciation|\\resumption \rightarrow citizen$
\end{itemize}

Applying these rules, we can integrate different call centre data into a combined internal data set for deeper analysis.
\subsection{Time Series Analysis}
\subsubsection{Time Series Transformation}
\label{tsTransformation}
The two attributes ``Disposition Code" and ``Disposition Code Text" are significant for investigating the correlations for specific topics, because the attributes are supervised and evaluated by the domain experts. If we transform the internal call centre data by time series using frequency and percentage strategy as shown in Eqn. \ref{eqn-frequency} and \ref{eqn-percentage}, the interaction records in Table \ref{tab:toyExample_callCentre} could be transformed to Table \ref{tab:toyExample_TS_callCentre}.

\begin{equation}
frequency(c_i,t_i) = num(c_i,t_i)
\label{eqn-frequency}
\end{equation}

\begin{equation}
percentage(c_i,t_i) = \frac{num(c_i,t_i)}{num(t_i)}
\label{eqn-percentage}
\end{equation}
where, $num(c_i,t_i)$ is the volume number of the queries for specific topic $c_i$ on time $t_i$, and $num(t_i)$ is the number of all the enquiries on $t_i$.

\begin{table}[htbp]
  \centering
  \caption{Time Series for Call Centre}
    \begin{tabular}{|c|c|c|c|}
      \hline
      Date & Topic & Frequency & Percentage \\
      6/12/2013 & 457 visa & 2 & 2/3 \\
      6/12/2013 & student visa & 1 & 1/3 \\
      7/12/2013 & 600 visit & 2 & 2/3 \\
      8/12/2013 & skilled migration & 1 & 1/3 \\
      \hline
    \end{tabular}
\label{tab:toyExample_TS_callCentre}
\end{table}

\begin{figure}
\centerline{\includegraphics[scale=1.0]{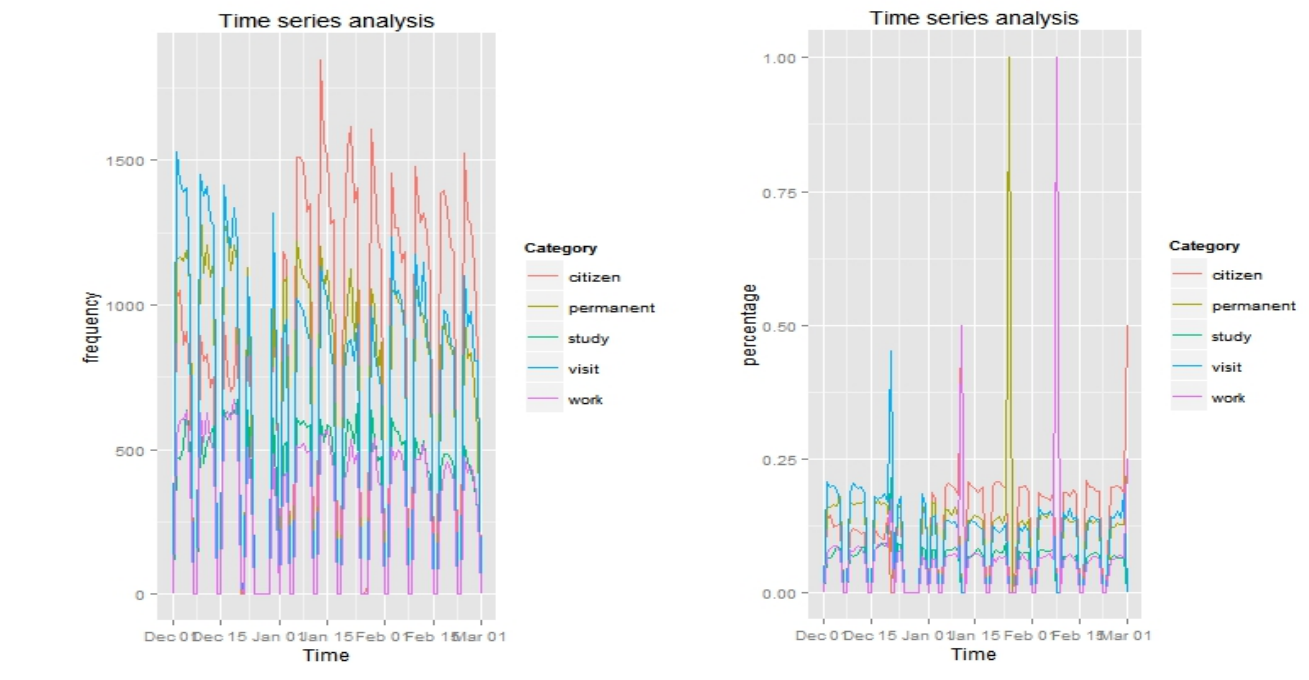}}
\caption{Frequency and Percentage Strategy for Time Series Analysis}
\label{fig_timeseries}
\end{figure}

Through the above two strategies, the call centre data can be transformed into topical time series. Fig. \ref{fig_timeseries} shows the transformed topical time series respectively for frequency and percentage strategies. To analyse the couplings between call centre and twitter, we also need to convert the public media data into the similar topical time series. Keyword searching is the basic method for this transformation, the main steps include:
\begin{itemize}
  \item Based on the extracted twitter data, we first apply keyword searching to get specific topical data. For instance, the data regarding ``student visa" topic could be collected by combining searching results of  ``student visa", ``560", ``570", ``571", ``572", ``573", ``574", ``575", and ``576".
  \item For specific topic, we respectively apply the above frequency and percentage strategies to transform the searched data into topical time series.
\end{itemize}

\subsubsection{Time Series Decomposition}
From the combined call centre data, basically we can get five groups of time series regarding different topics. From the call centre data of DIBP, there are no enquires during weekends and public holidays because call centre is not open during that time, causing seasonal sharp troughs for the time series changing trends. This seasonal effect is a systematic and calendar related effect caused by public holidays. To better understand the changing patterns of the time series, the observed data needs to be seasonally adjusted as seasonal effects can conceal both the true underlying movement in the series, as well as certain non-seasonal characteristics which may be of interest to analysts and business. Thus, the time series decomposition method is applied to remove the seasonal effect on time series.

It is well known that time series can be decomposed into three components: the trend (long term direction), the seasonal and the irregular (unsystematic, short term fluctuations). The trend is defined as the `long term' movement in a time series without calendar related and irregular effects, and is a reflection of the underlying level. The seasonal component consists of effects that are reasonably stable with respect to timing, direction and magnitude. It arises from systematic, calendar related influences such as: natural weather conditions; business or administrative procedures; social and cultural behaviour such as Christmas and Chines New Year. The irregular component is what remains after the seasonal and trend components of a time series have been estimated and removed. It results from short term fluctuations in the series which are neither systematic nor predictable.

\begin{figure}
\centerline{\includegraphics[scale=0.5]{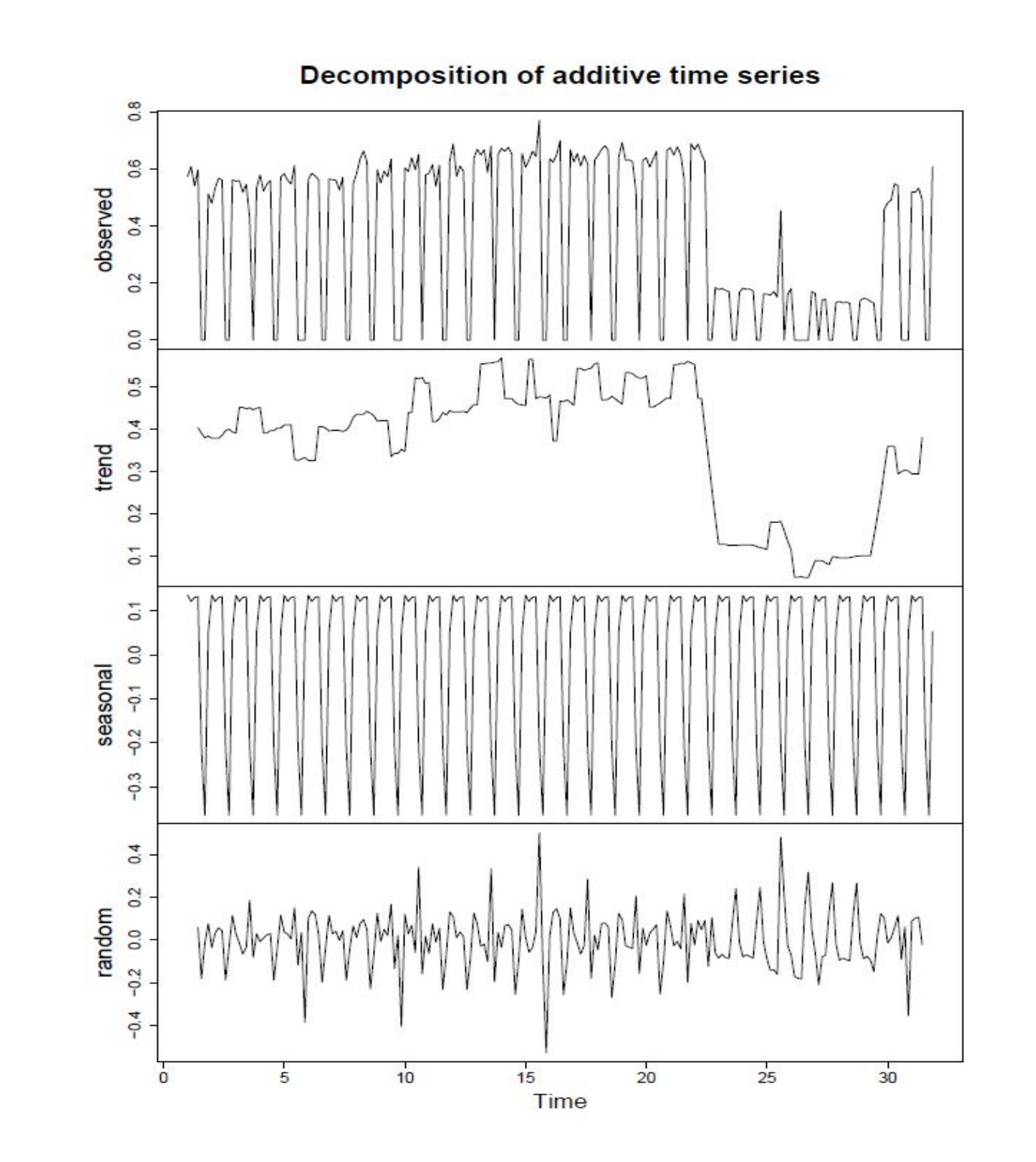}}
\caption{Example of Time Series Decomposition}
\label{fig_tsDecomposition_example}
\end{figure}

\begin{figure}
\centerline{\includegraphics[scale=1.0]{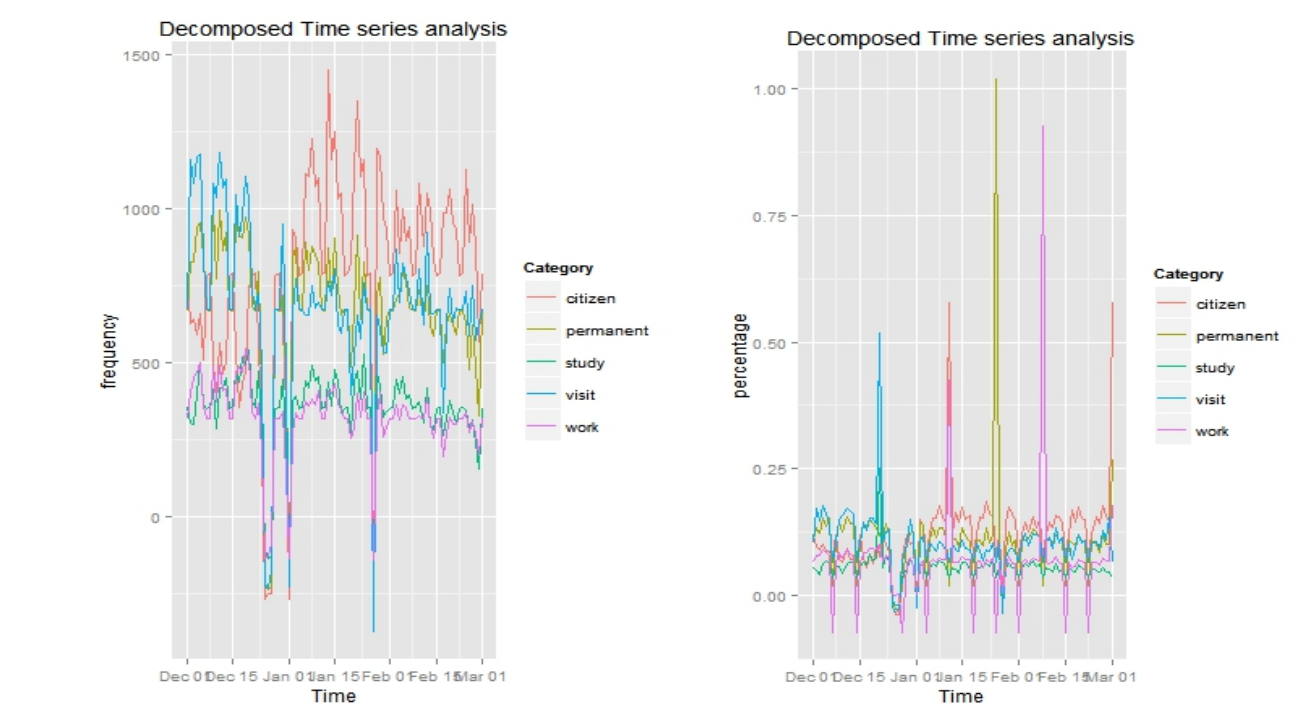}}
\caption{Frequency and Percentage Strategy for Decomposed Time Series Analysis}
\label{fig_decomposed_timeseries}
\end{figure}

To decompose the call centre time series data, we apply the ``decompose" function by R language using moving averages with addictive model. Specifically, this function first determines the trend component using a moving average with equal weighted symmetric window, and removes it from the time series. Then, the seasonal figure is computed by averaging method over all periods for each time unit. The seasonal figure is then centered. Finally, the irregular component is determined by removing trend and seasonal figure from the original time series. Fig. \ref{fig_tsDecomposition_example} shows an example of the time series decomposition for ``visit" category. Fig. \ref{fig_decomposed_timeseries} shows the decomposed time series based on frequency and percentage strategies.

\subsection{Correlation Analysis}
\subsubsection{Cross-Correlation Analysis}
In time series analysis, as applied in statistics and signal processing, the cross correlation between two time series describes the normalized cross covariance function. The cross correlation of two time series can be estimated by averaging the product of samples measured from one process and samples measured from the other time series. The samples included in the average can be an arbitrary subset of all the samples in the time series. For a large number of samples, the average converges to the true cross-correlation.

Let $(x_t,y_t)$ represent two time series, then the cross correlation is given by Eqn. \ref{eqn-cross_correlation}:
\begin{equation}
Corr_{xy}(\tau) = E[(x_t-\mu_x)(y_{t+\tau}-\mu_y)]
\label{eqn-cross_correlation}
\end{equation}
where $\mu_x$ and $\mu_y$ are the means of $x_t$ and $y_t$ respectively.

\subsubsection{Lagged Correlation}

Lagged correlation refers to the correlation between two time series shifted in time relative to one another. Lagged correlation is important for studying the couplings between time series for two reasons. First, one series may have a delayed response to the other series, or perhaps a delayed response to a common stimulus that affects both series. For example, some people may first access social media to solve their problems because of the quick response of social media. If they cannot find the answer, they may choose to call the authorities regarding the problem. Second, the response of one series to the other series or an outside stimulus may be ¡°smeared¡± in time, such that a stimulus restricted to one observation elicits a response at multiple observations.

The simple correlation coefficient between two time series properly aligned in time is inadequate to analyse the couplings. A very useful function for checking lagged correlations is the sample cross correlation function (CCF). The CCF function is the correlation between the series shifted against one another as a function of number of observations of the offset. If the individual series are autocorrelated, the estimated cross-correlation function may be distorted and misleading as a measure of the lagged relationship.

In the relationship between two time series ($y_t$ and $x_t$), the series $y_t$ may be related to past lags of the $x$-series. The CCF function is helpful for identifying lags of the $x$-variable that might be useful predictors of $y_t$. In this paper, we use this CCF function defined as the set of sample correlations between $x_{t+h}$ and $y_t$ for $h=0,\pm 1,\pm2,\pm3 ...$ to determine the time lags between call centre and twitter. A negative value for $h$ is a correlation between the $x$-variable at a time before $t$ and the $y$-variable at time $t$. For instance, consider $h=-2$, the CCF value would give the correlation between $x_{t-2}$ and $y_t$. Simply, the following two rules are applied to identify whether one time series leads or lags the other time series.

\begin{itemize}
  \item When one or more $x_{t+h}$, with $h$ negative, are predictors of $y_t$, it is said that $x$ leads $y$.
  \item When one or more $x_{t+h}$, with $h$ positive, are predictors of $y_t$, it is called that $x$ lags $y$.
\end{itemize}

According to the above rules, we can set different positive value for $h$ to examine the delayed time between time series $x$ and time series $y$. After calculating the cross-correlation between the two time series, the maximum (or minimum if the two series are negatively correlated) of the cross-correlation function indicates the point in time where the time series are best aligned, i.e. the time delay between the two series is determined by the argument of the maximum of the cross-correlation as in Eqn. \ref{eqn-lagged_correlation}

\begin{equation}
\tau_{delay} = \mathop {\arg \max }\limits_h \left( {Corr\left( {{x_{t + h}},{y_t}} \right)} \right)
\label{eqn-lagged_correlation}
\end{equation}

\subsubsection{Time Series Symbolic Representation}

Through the time series decomposition method, we can partly remove the seasonal patterns. However, if we look at the decomposed time series for call centre data as shown in Fig. \ref{fig_decomposed_timeseries}, it can be clearly seen that there is still some big decrease during weekends and public holidays, probably because the seasonal patterns are not completely removed. Therefore, we apply a symbolic aggregate approximation (SAX)\cite{Lin:2007:ESN:1285960.1285965} method to transform the real valued time series into symbolic representations. SAX is the first symbolic representation for time series that allows for dimensionality reduction and indexing with a lower-bounding distance measure. Symbolic Aggregate approximation was proposed by Lin et al. and extends piecewise aggregate approximation (PAA) based approach\cite {Keogh:2005:EID:1047750.1047753} inheriting algorithm simplicity and low computational complexity while providing satisfiable sensitivity and selectivity in range-query processing.

SAX transforms a time-series $x$ of length $n$ into the string of arbitrary length $\omega$, where $\omega \leq n$  typically, using an alphabet $A$ of size $a > 2$. The SAX algorithm consist of two steps: during the first step it transforms the original time-series into a PAA representation and this intermediate representation gets converted into a string during the second step. Use of PAA at the first step brings the advantage of a simple and efficient dimensionality reduction while providing the important lower bounding property. The second step, actual conversion of PAA coefficients into symbols, is also computationally efficient.

To better analyse the correlations between call centre and public media, we follow this direction and first transform the original time series into a PAA representation and then apply the symbolic representation of the changing trends to help understand the correlations.

\subsubsection{Correlation Analysis Interface}	

\begin{figure}
\centerline{\includegraphics[scale=1.0]{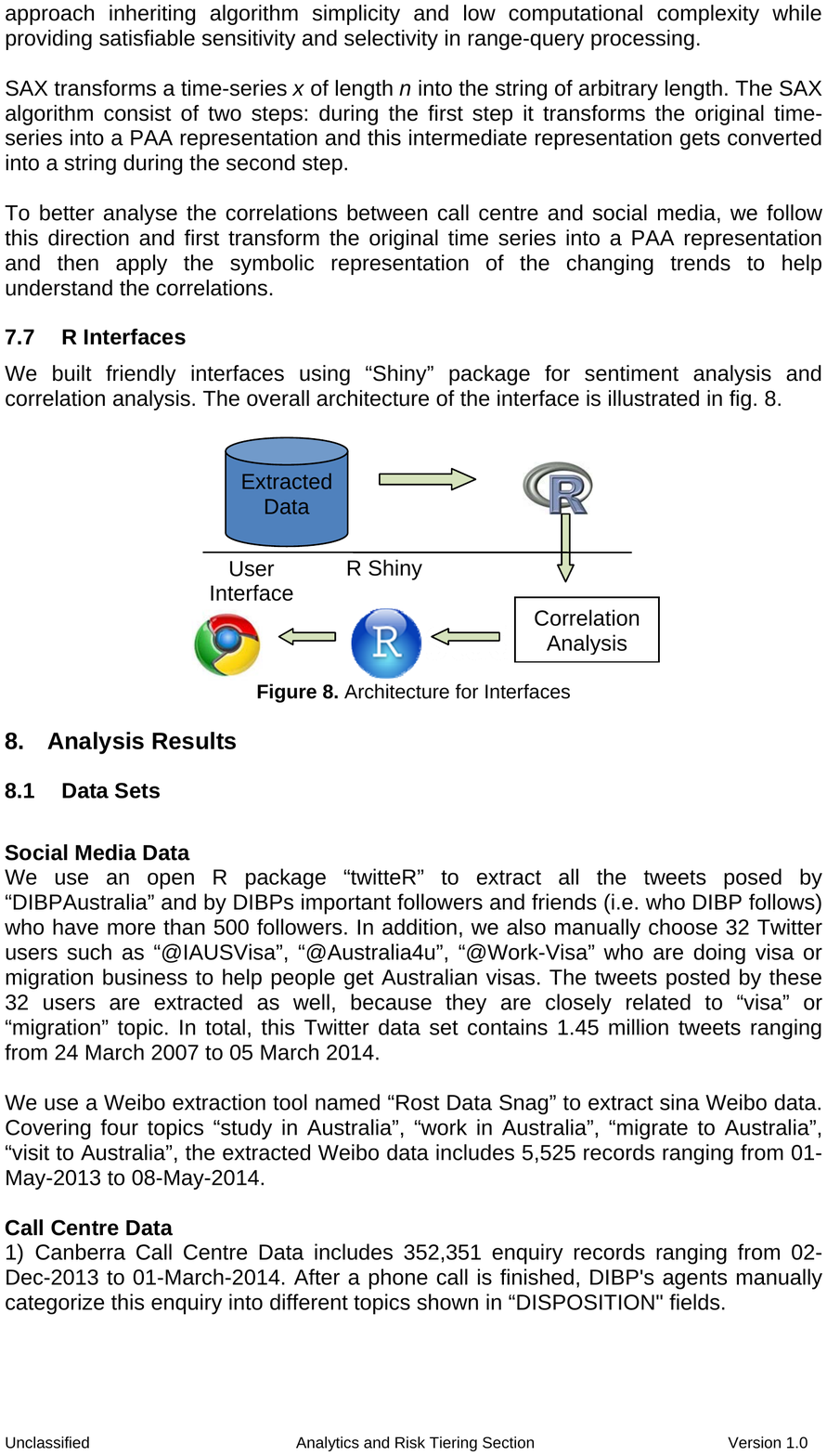}}
\caption{Architecture of the Interface for Correlation Analysis}
\label{fig_interface_framework}
\end{figure}

We implemented a friendly interface with the architecture in Fig. \ref{fig_interface_framework} for correlation analysis between call centre and social media. Users can simply select one category to see the correlations during different time periods. Then the cross-correlations and time lagged correlations by different strategies can be clearly seen from the interface. For Twitter and call centre, the interface includes eight categories: ¡°study¡±, ¡°visit¡±, ¡°work¡±, ¡°permanent¡±, ¡°citizen¡±, ¡°refugee¡±, ¡°skilled permanent¡±, and ¡°working holidays¡±.

\section{Experimental Analysis}
The experimental steps are as follow: we first extract external twitter data set, and integrate call centre data and twitter data together, then apply time series transformation introduced in Section \ref{tsTransformation} to transform the data sets into topical time series. After that, we use different correlation analysis methods to study the couplings between call centre and twitter.

\subsection{Data Set}
\subsubsection{Social Media Data}

We use an open R package ``twitteR" to extract all the tweets posed by ``DIBPAustralia" and by DIBP's important followers and friends who have more than 500 followers. In addition, we also manually choose 32 twitter users such as ``@IAUSVisa", ``@Australia4u", ``@Work-Visa" who are doing visa or migration business to help people get Australian visas. The tweets posted by these 32 users are extracted as well, because they are closely related to ``visa" or ``migration" topic. In total, this extracted twitter data has 1.45 million tweets ranging from 24 March 2007 to 05 March 2014.

\begin{figure*}
\centerline{\includegraphics[scale=0.6]{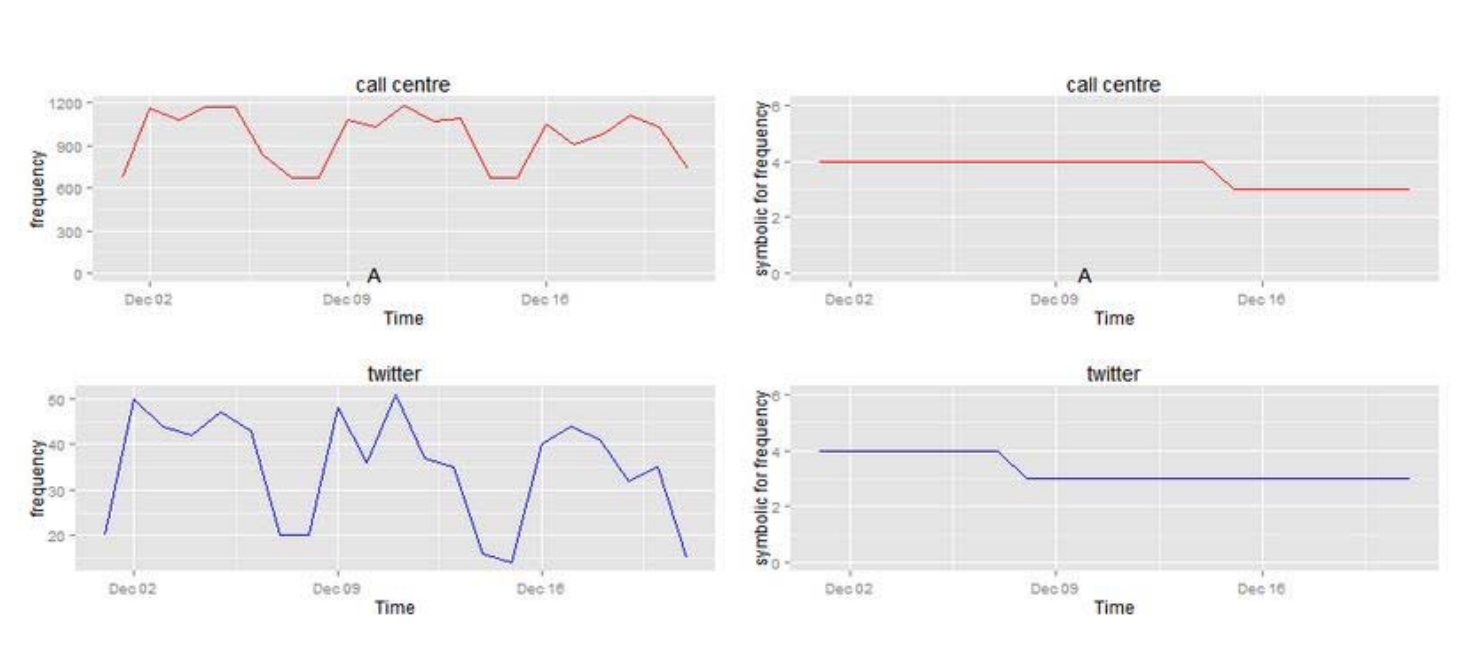}}
\caption{Frequency Strategy for Visit Category}
\label{fig_ex_fig1_frequency}
\end{figure*}

\begin{figure*}
\centerline{\includegraphics[scale=0.6]{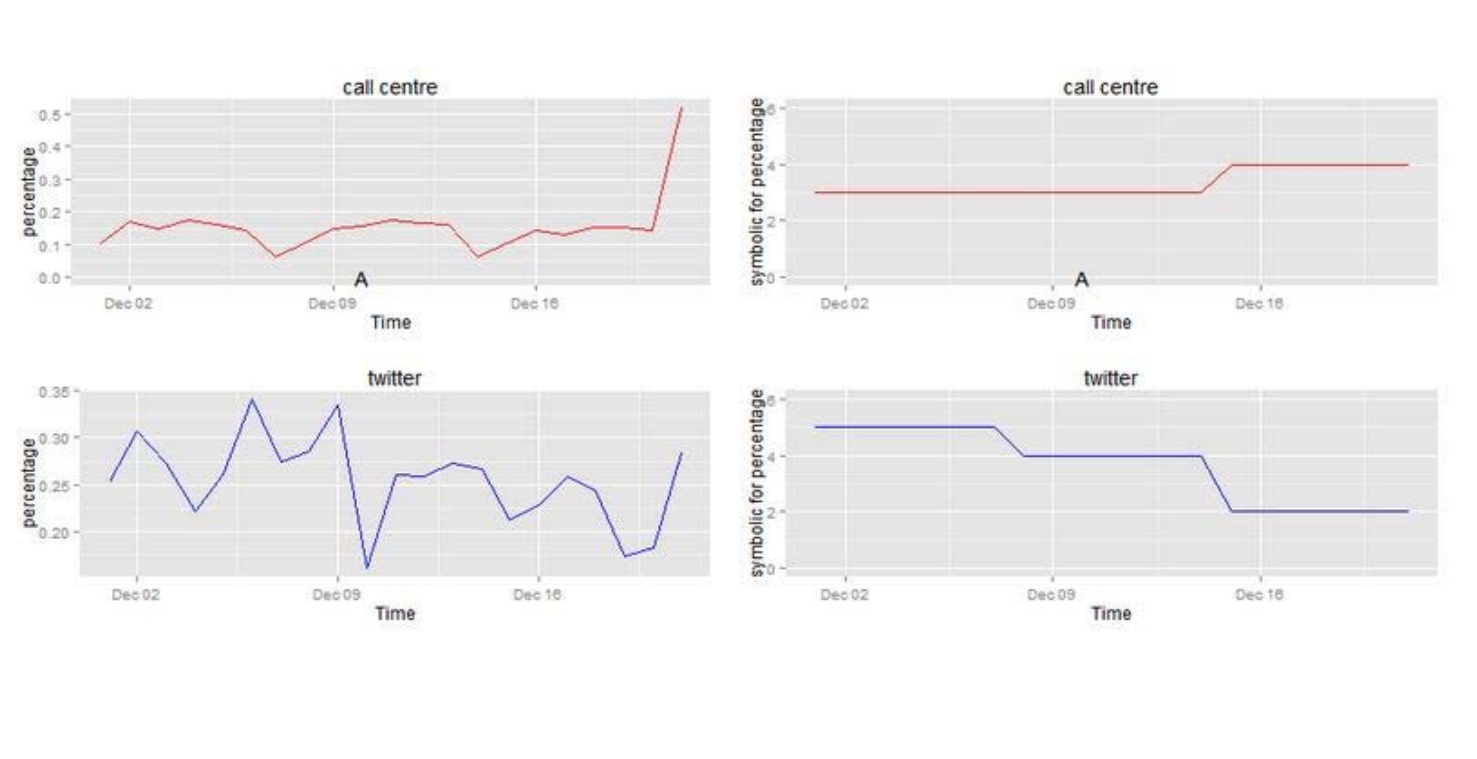}}
\caption{Percentage Strategy for Visit Category}
\label{fig_ex_fig1_percentage}
\end{figure*}

\subsubsection {Call Centre Data}
In this paper, we combine DIBP's three different call centre data respectively from Canberra, Ottawa, and London. The details of the three call centre data are as follow:
\begin{itemize}
\item Canberra Call Centre Data includes 352,351 enquiry records ranging from 02-Dec-2013 to 01-March-2014. After a phone call is finished, DIBP's call centre agents manually categorize this enquiry into different topics shown in ``DISPOSITION" fields.
\item Ottawa Call Centre Data contains 103,069 records (emails and phone calls) ranging from 2-July-2013 to 12-March-2014. Field ``Category" shows the topics of the enquires.
\item London Call Centre Data includes 175,450 records ranging from 01-Jan-2013 to 28-Feb-2014.
\end{itemize}

Since the three call centre data are from different countries and have different labels to identify topic information, we integrate these call centre data together by the merging rules introduced in Section \ref{callCentreIntegration}. After this integration, we get the completely overlapped combined call centre data including 387,037 records ranging from 01-Dec-2013 to 28-Feb-2014.

\subsection{Typical Discovered Correlations}
%

In visit category as shown in fig. \ref{fig_ex_fig1_frequency} and \ref{fig_ex_fig1_percentage} during 01 Dec, 2013 to 21 Dec, 2013, no delay with the probability 0.86 for frequency strategy, and twitter lags call centre one day with the correlation value -0.44 for percentage strategy. The results of the symbolic representation also show that frequency-based trends positively correlated and percentage-based trends negatively correlated.

\begin{figure}
\centerline{\includegraphics[scale=0.6]{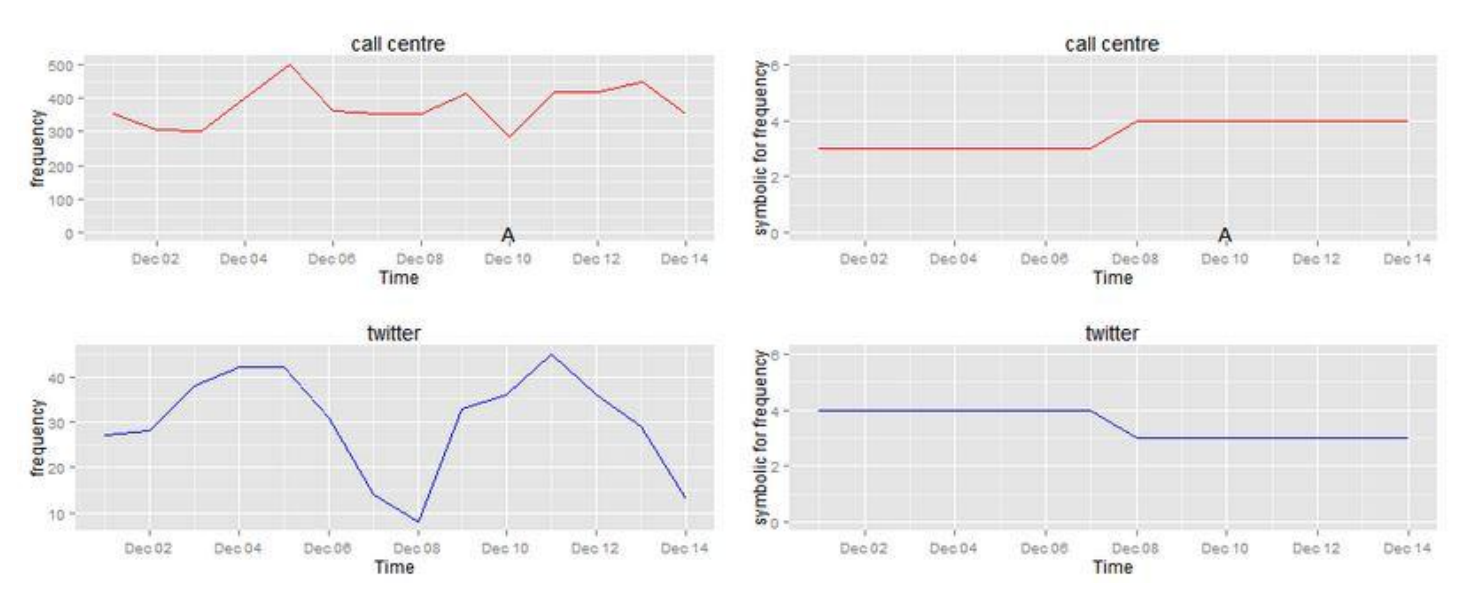}}
\caption{Frequency Strategy for Study Category}
\label{fig_ex_fig2_frequency}
\end{figure}

\begin{figure}
\centerline{\includegraphics[scale=0.6]{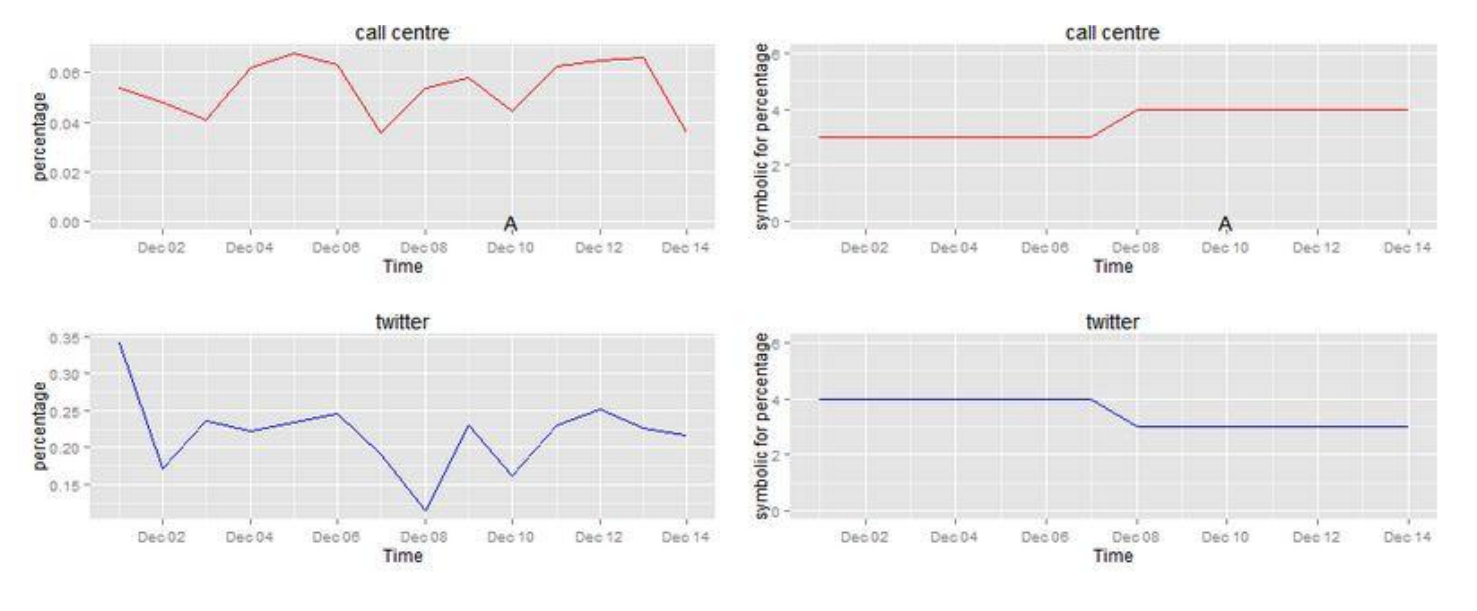}}
\caption{Percentage Strategy for Study Category}
\label{fig_ex_fig2_percentage}
\end{figure}

In study category as shown in fig. \ref{fig_ex_fig2_frequency} and \ref{fig_ex_fig2_percentage}, during 01 Dec, 2013 to 14 Dec, 2013, we observe that call centre delays three days with the correlation -0.58 for frequency strategy, but no delay using percentage strategy with the similarity 0.29. From the symbolic representation, the results show that the two changing trends are negatively correlated during this period.

%
%
%
%
%

\begin{figure}
\centerline{\includegraphics[scale=0.6]{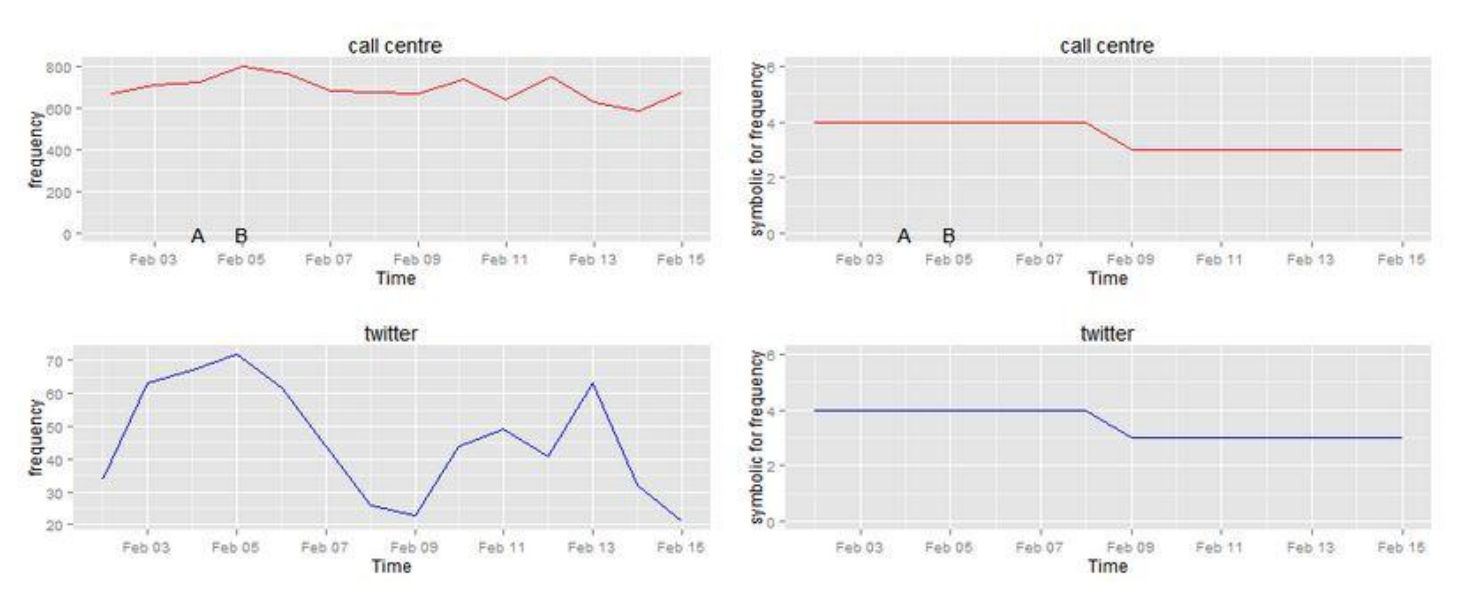}}
\caption{Frequency Strategy for Permanent Category}
\label{fig_ex_fig5_frequency}
\end{figure}

\begin{figure}
\centerline{\includegraphics[scale=0.6]{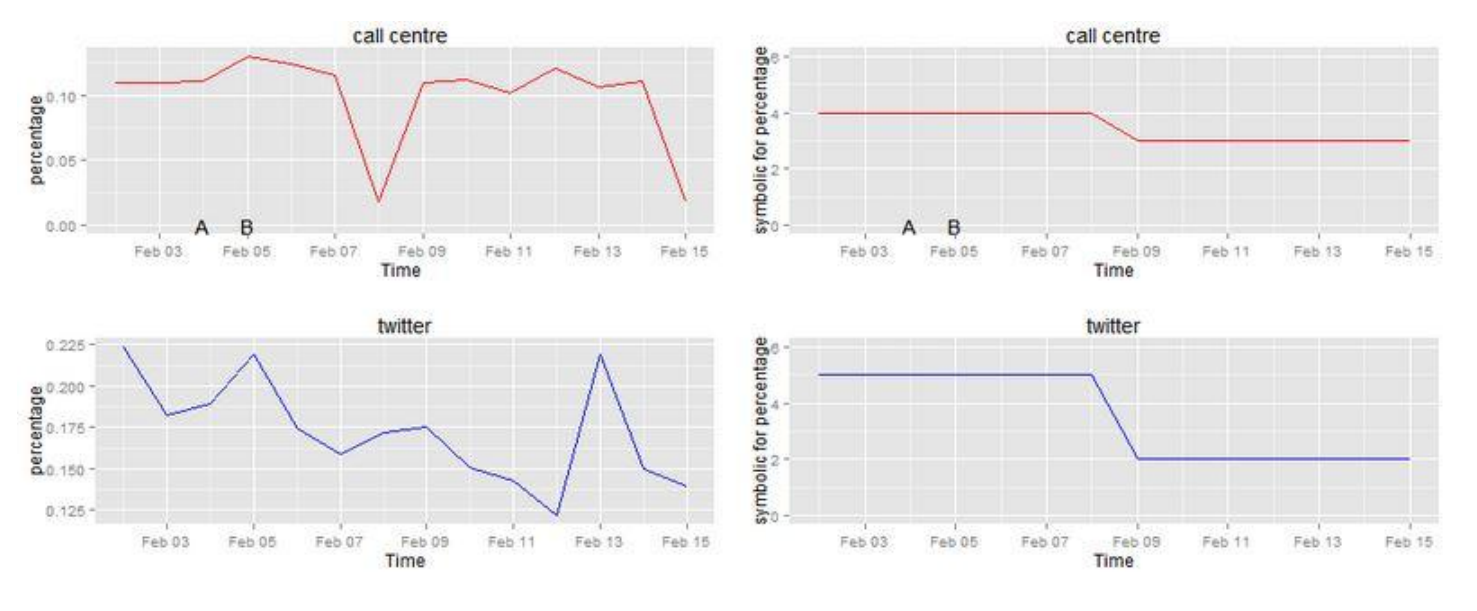}}
\caption{Percentage Strategy for Permanent Category}
\label{fig_ex_fig5_percentage}
\end{figure}

In permanent category as shown in fig. \ref{fig_ex_fig5_frequency} and \ref{fig_ex_fig5_percentage}, during 02 Feb, 2014 to 15 Feb, 2014, call centre lags one day for frequency strategy with the probability 0.62, while lags five days using percentage strategy with the similarity -0.46. The weekly aggregated results of symbolic representation have the same changing trends.

\begin{figure}
\centerline{\includegraphics[scale=0.6]{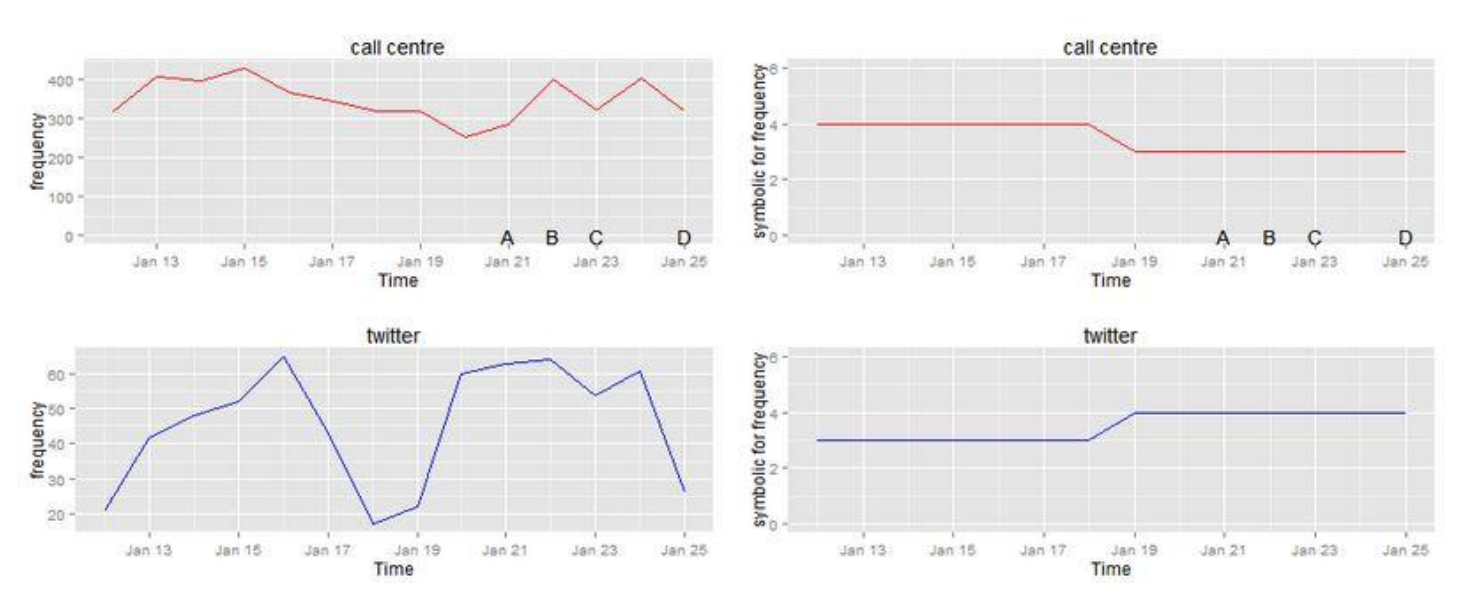}}
\caption{Frequency Strategy for Work Category}
\label{fig_ex_fig6_frequency}
\end{figure}

\begin{figure}
\centerline{\includegraphics[scale=0.6]{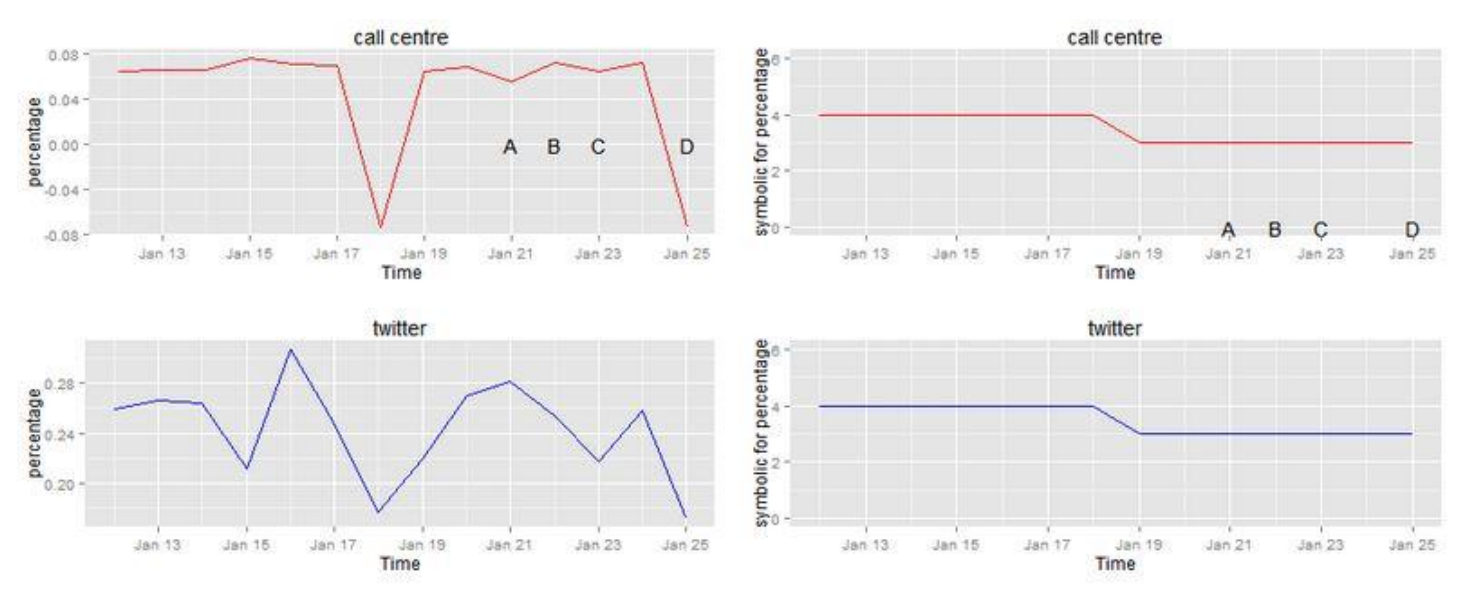}}
\caption{Percentage Strategy for Work Category}
\label{fig_ex_fig6_percentage}
\end{figure}

In work category as shown in fig. \ref{fig_ex_fig6_frequency} and \ref{fig_ex_fig6_percentage}, during 12 Jan to 25 Jan, there is no delay for call centre and twitter with probability 0.74 for percentage strategy. However, call centre delays three days with -0.5 for frequency strategy. The symbolic results also demonstrate the same changing trends that frequency-based trends are negatively correlated, but percentage-based trends are positively correlated.

\begin{figure}
\centerline{\includegraphics[scale=0.6]{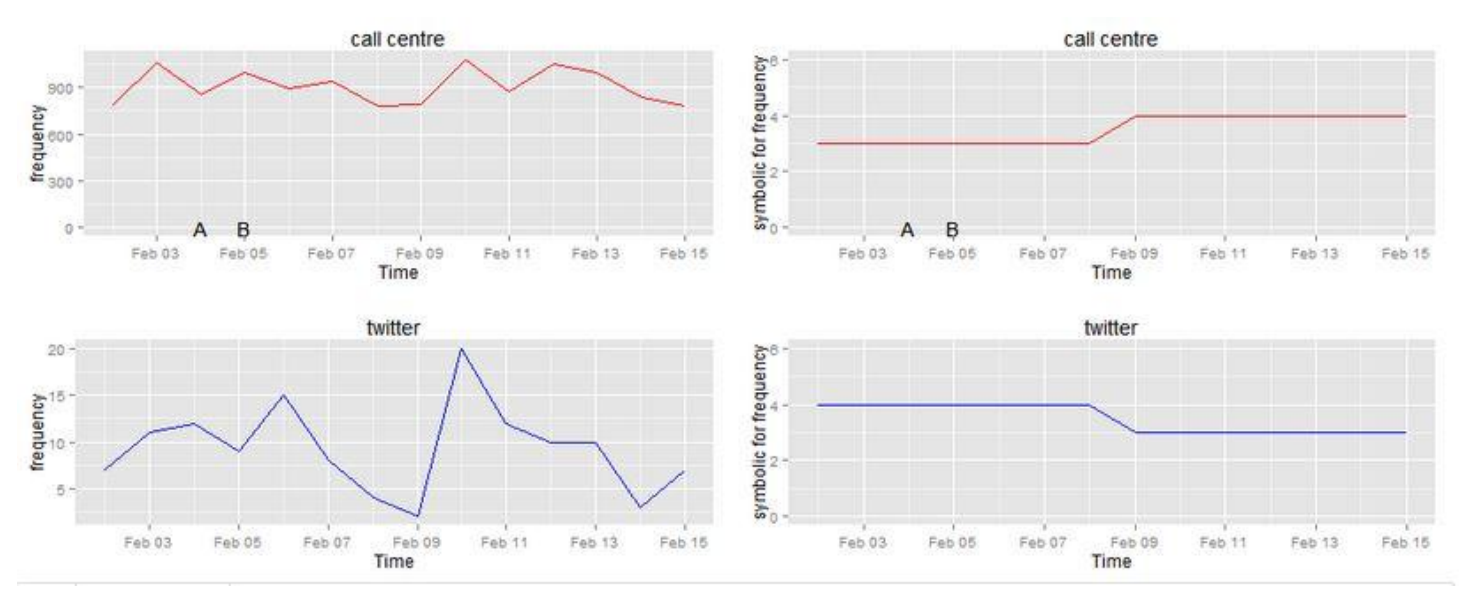}}
\caption{Frequency Strategy for Citizen Category}
\label{fig_ex_fig7_frequency}
\end{figure}

\begin{figure}
\centerline{\includegraphics[scale=0.6]{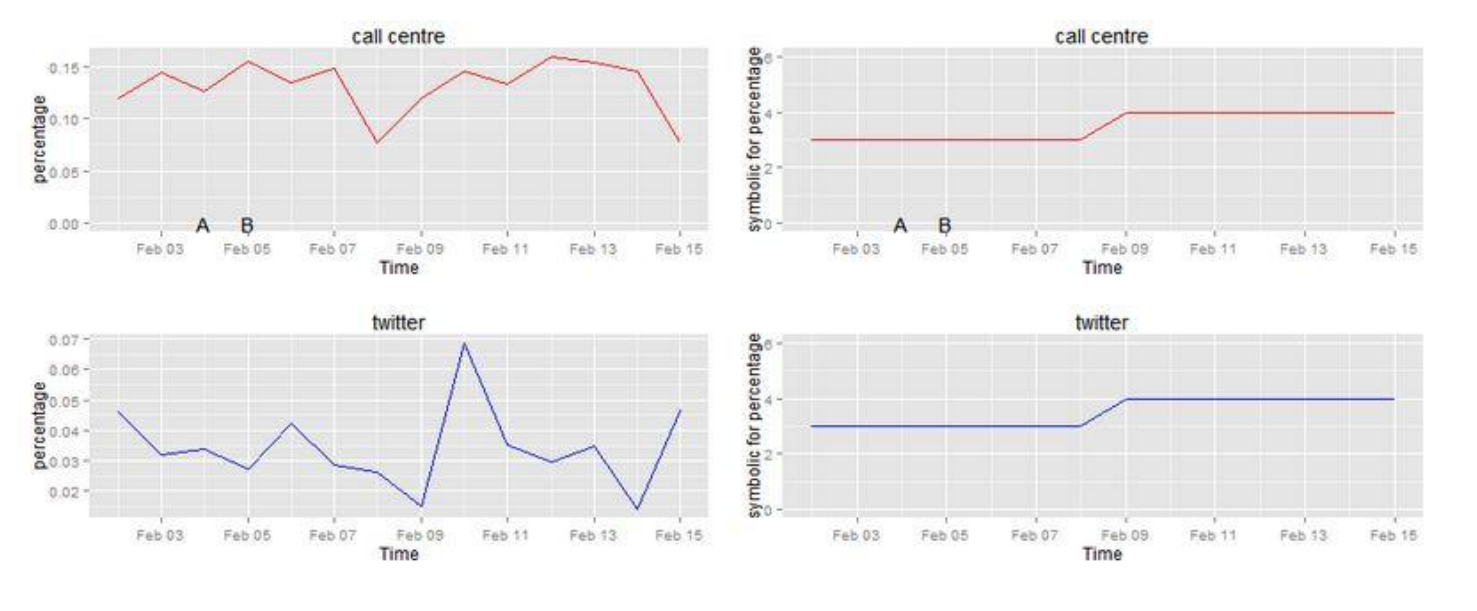}}
\caption{Percentage Strategy for Citizen Category}
\label{fig_ex_fig7_percentage}
\end{figure}

In citizen category as shown in fig. \ref{fig_ex_fig7_frequency} and \ref{fig_ex_fig7_percentage}, during 02 Feb to 15 Feb, no delay with the correlation value 0.65 for frequency strategy, but call centre delays two days with -0.61 for percentage strategy. Differently, the symbolic representation results indicate that frequency-based trends are negatively correlated, and percentage-based trends are positively correlated.

\subsection{Discussions}
From the above discovered couplings between call centre and twitter, we can clearly see that the two data sources indeed have some coupling relations and time lagged correlations. The time lagged correlation can be concluded as three situations: 1) call centre lags twitter is the most common case, which means that social media normally has a quicker response regarding different topics, and people will turn to call centre to verify or confirm the discussed information on twitter. 2) call centre and twitter has no delay means that some events have big influence so that the discussions regarding the topic on twitter and call centre both increase quickly. 3) twitter lags call centre is not very common but indeed happen in our discovered couplings, which represent that some people prefer to directly communicate with DIBP to get the answers, then to share or discuss further information with their friends on social media.

It is still difficult to determine whether the correlations are influenced by some relevant events or policies related to DIBP, because the significant events themselves are not easy to be confirmed. But the positive response from business staff in DIBP can partly demonstrate that this coupling learning between call centre and twitter is very interesting and meaning.

\section{Conclusion and Future Work}
In this paper, we studied the coupling analysis between DIBP's call centre and external twitter data, and proposed a coupling analysis framework. First, we transformed the extracted twitter and call centre data to time series, then applied time series decomposition to remove seasonal patterns. After that, time series correlation analysis with frequency and percentage strategies was used to discover the couplings. Besides this, to better understand the couplings, we also applied the symbolic representation method to show the couplings. The experimental analysis parts demonstrated that the two data sources indeed have coupling relations, which are significant for business.

The call centre data currently only covers three months from December 2013 to February 2014, which is not enough to get a convincible analysis results. To continue more effective analysis, it is necessary to collect more call centre data for overcoming the limitations of data sets.

Regarding the correlation analysis part, currently we just analysed the overall changing trends and their correlations on call centre and social media. We actually found that correlations indeed exist between the two data sources. However, the events or significant policies which incur the correlations and the changing trends are still not clear, because just very rare policies have been published on social media websites, which cause tough to find the related policies. More significant policies could be beneficial for deeper understanding of correlations.

\section{Acknowledgments}
This work is sponsored in part by the joint project between University of Technology Sydney and Australian Department of Immigration and Border Protection, Australian Research Council Discovery Grants (DP1096218 and DP130102691) and ARC Linkage Grant (LP100200774).
\bibliography{ref}
\bibliographystyle{plain}

\section{Biography}

\subsection{Fangfang Li}
Dr. Fangfang Li currently is a PhD candidate in the Advanced Analytics Institute at University of Technology Sydney. He received his first PhD degree in Computer Science from Beijing Institute of Technology, China. Now he is working toward the second PhD degree of data analytics in University of Technology Sydney.

\subsection{Yanchang Zhao}
Dr. Yanchang Zhao is a founder of RDataMining.com and an RDataMining Group on LinkedIn, and an organizer of the Canberra Data Miners Group. He has been a Senior Data Miner in Australian public sector since 2009. Before joining public sector, he was an Australian Postdoctoral Fellow (Industry) at University of Technology, Sydney, Australia. He started his research on data mining since 2001 and has been applying data mining in in Australia public sector since 2006. His current research interest is using R for data mining applications. He has 50+ publications on data mining research and applications, including three books. He has been a Program Chair of the Australasian Data Mining Conference (AusDM 2012-2014), a Program Chair of the 2013 International Workshop on Data Mining Applications in Industry and Government (DMApps 2013), and a program committee member for more than 60 international conferences. He is a senior member of the IEEE and a member of the Institute of Analytics Professionals of Australia.

\subsection{Klaus Felsche}
Klaus Felsche is the director of Analytics \& risk Tiering at Department of Immigration and Border Protection, Australia. He has broad leadership experience in education, training, capability development, policy formulation, intelligence, IT systems, project leadership and delivery of business outcomes enabled by processes that encourage innovation. His specialty includes deployment of advanced analytics in a transaction-based environment and policy delivery.

\subsection{Guandong Xu}
Dr. Guandong Xu currently is a lecturer in the Advanced Analytics Institute at University of Technology Sydney following a research fellow in Centre for Applied Informatics at Victoria University. He received MSc and BSc degree in Computer Science and Engineering from Zhejiang University, China. He gained PhD degree in Computer Science from Victoria University. After that he worked as a Postdoctoral research fellow in the Centre for Applied Informatics at Victoria University and then Postdoc in Department of Computer Science at Aalborg University, Denmark. He was an Endeavour Postdoctoral Research Fellow in the University of Tokyo in 2008.

\subsection{Longbing Cao}
Prof. Longbing Cao got one PhD in Intelligent Sciences in Chinese Academy of Sciences and another in Computing Science from the University of Technology Sydney (UTS). He is a professor of information technology at UTS, the Founding Director of the university's research institute Advanced Analytics Institute and a core member of the Data Sciences and Knowledge Discovery Lab at the Centre for Quantum Computation and Intelligent Systems at the Faculty of Engineering and IT, UTS. He is also the Research Leader of the Data Mining Program at the Australian Capital Markets Cooperative Research Centre, the Chair of ACM SIGKDD Australia and New Zealand Chapter, IEEE Task Force on Data Science and Advanced Analytics, and IEEE Task Force on Behavioral, Economic and Socio-cultural Computing. He is a Senior Member of IEEE, SMC Society and Computer Society. He serves as associate editor and guest editor on such journals as
ACM Computing Surveys, as conference co-chair of KDD2015, PAKDD13 and ADMA13, and program co-chair or vice-chair of PAKDD11, ICDM10, IAT11, ADMA10 etc., and program committee member on around 100 conferences including KDD, ICDM and AAMAS. He joined UTS in 2005 after he served as an editor, marketing strategist and then Chief Technology Officer in China.

\end{document}